\documentclass[twocolumn,showpacs,preprintnumbers,amsmath,amssymb,floatfix]{revtex4}

\usepackage{graphicx}
\usepackage{dcolumn}
\usepackage{bm}

\begin{document}

\title{Phase separation and electron pairing in repulsive Hubbard clusters}

\author{G. W. Fernando$^{1,4}$,  
 A. N. Kocharian$^2$, K. Palandage$^1$, Tun Wang$^1$, and J. W. Davenport$^3$}
\address{$^1$U-46, Physics Dept., University of Connecticut,
Storrs, CT 06269}
\address{$^2$Physics Dept.,  California State University, Northridge, CA 91330}
\address{$^3$Physics Dept., Brookhaven National Laboratory, 
Upton, NY 11973}
\address {$^4$Institute of Fundamental Studies, Hantana Road, Kandy, Sri Lanka}

\begin{abstract}

  Exact thermal studies of small (4-site, 5-site and 8-site)
 Hubbard clusters with local electron repulsion
 yield intriguing insight
 into  phase separation,
  charge-spin separation, pseudogaps, condensation,
 in particular, pairing fluctuations away from half filling
 (near optimal doping).
 These exact calculations, carried out in
 canonical (i.e. for fixed electron number $N$)
 and grand canonical
(i.e. fixed chemical potential $\mu$) ensembles,
 monitoring variations in temperature
$T$ and magnetic field $h$, show rich phase diagrams in a
 $T$-$\mu$ space
 consisting of pairing fluctuations and signatures of
 condensation.  These electron pairing instabilities are seen
 when the onsite Coulomb interaction $U$ is
 smaller than a critical value $U_c(T)$ and they point to a possible
 electron pairing mechanism.
   The specific
 heat, magnetization, charge pairing and spin pairing
 provide strong support for the existence of competing
 (paired and unpaired) phases near optimal doping in
  these clusters as observed in recent experiments
in doped La$_{2-x}$Sr$_x$CuO$_{4+y}$ high  T$_c$
superconductors.

\end{abstract}

\maketitle
 Since the discovery of the high temperature superconductors (HTSCs),
 there has been an intense debate about a possible electron
 (or hole) pairing mechanism. Early on,
 P.~W.~Anderson~\cite{RVB}
 suggested
 that the large positive onsite Coulomb interaction in the Hubbard model
 should contain the key to some of the perplexing physics observed in the
 HTSCs. Although it is next to impossible to list every single
 effort related to testing the above assertion, important progress has
 been made in attempts to obtain a better
 understanding of the physical
 properties of these materials~\cite{Anderson,Nature,Timusk,Marshall,
 Kivelson_Review, Andrea}.
The bad metallic behavior and  small correlation length of
dynamical spin fluctuations (different from conventional
superconductors) make  HTSCs~\cite{Zha} a suitable platform to examine
the role of local interactions.
Exact studies of the Hubbard clusters placed in magnetic field $h$
are fundamental for understanding the nature of magnetism and
corresponding spin gaps in various cluster
geometries~\cite{Pastor,Sebold}.
Repulsive interactions can lead  to  phase separation,
electron pairing and ground state superconductivity in  certain
mesoscopic structures~\cite{scalettar,white}.
In our opinion,
thermal properties of small clusters with strong correlations have
not been fully explored, although there have been numerous
 exact calculations~\cite{Shiba,schumann}, and the present study is an
 attempt to fill this void.

  Our recent work~\cite{PRB} indicates that
  Hubbard clusters,
 when connected to a particle
 reservoir and a thermal bath,
 possess a vivid variety
 of interesting thermal and
 physical properties, that could pave the way for a new class
 of tunable materials. These inferences were drawn by not only
 carrying out exact diagonalizations of the many-body Hamiltonian,
 but also  using these eigenvalues
 in a statistical ensemble to study thermal
  and other transitions,
 by monitoring, for example, susceptibilities, {\sl i.e. fluctuations}.
 The many-body nature of these correlated problems is at least
partly hidden in statistical fluctuations and it is no wonder
that these fluctuations give rise to intriguing results. The
crossovers and transitions between various phases,
 that we identify at finite and zero temperatures,
 are found by monitoring the corresponding
 thermal and ground state properties
 without taking the thermodynamic limit. The 
 results described in this paper provide new 
 insights into the physics of the 4-site as well as larger (repulsive)
 Hubbard clusters.

           These attempts may be questioned
  since they appear not to comply with the standard
  applications of
 statistical mechanics with respect to the thermodynamic limit.
  However, for finite
  systems, it is necessary to re-evaluate these ideas
  and a paradigm shift in our thinking
 may be necessary.
 We have already shown
 that in such finite systems, one can define and identify
 various transitions and phase
 boundaries by monitoring maxima and minima
 in susceptibilities~\cite{PRB}.
 As synthesis techniques improve at a
 rapid rate, it has become possible to synthesize isolated clusters and
 hence it is  clear that we need not
 always look at the thermodynamic limit.
 Finite, mesoscopic
 structures (i.e. clusters containing a few atoms) in suitable
 topological forms will be realistic enough to
 synthesize and extract fascinating
 physical properties.
  Also, since the HTSCs
  are known to consist of (stripes and
  possibly other) inhomogeneities ~\cite{Tranquada},
  it is possible that these cluster studies may be able to capture
  some of the essential physics of the HTSCs.
  The following is a list of properties, resulting from 
  our exact  (4-site, 5-site and 8-site) Hubbard cluster studies,
  that is shared with  the HTSCs.

\begin{itemize}
  \item  Phase diagrams in a temperature-chemical potential (doping) plane
      and the presence of a multitude of fascinating phases, including 
      Mott-Hubbard like paramagnetic and
       antiferromagnetic phases~\cite{PRB}.

  \item   Vanishing of a charge gap at a critical set of parameters
       and thereby providing an effective
       attraction leading to onset of electron
       charge pairing (2e)
       at a critical temperature $T^P_{c}$.

 \item   Spin pairing at a lower temperature
 ($T^P_{s}$) and hence the formation
       of rigidly bound spin pairs in a narrow, critical region of
       doping.

 \item    Low temperature specific heat peak, reminiscent of the
       experimental,
       low temperature specific heat behavior in the HTSCs~\cite{Tallon}.

 \item   Temperature vs $U$
         phase diagram, indicating the pressure effect
      on the superconducting transition temperature as seen in recent
      experiments~\cite{Xiao}.

 \item  The presence of a dormant magnetic state, lurking in the above narrow,
      critical region of doping, that could be stabilized by either
      applying a magnetic field, going above the spin pairing temperature,
      or changing the chemical potential,
      as seen in a recent, notable  experiment~\cite{hashini}.

  \item The opening of a pseudogap above the pairing temperature,
        as observed
        in NMR experiments,
        in both hole and electron doped cuprates~\cite{nmr}.

  \item   Larger clusters with different topologies and
         higher dimensionality illustrating how
         the above properties get scaled with size.

\end{itemize}

 In what follows, we will  address the similarities
 listed above using the many-body eigenvalues
 of the Hubbard clusters ({\sl with energies measured in units
of $t$, the hopping parameter})  in combination with
statistical mechanics.
 The grand partition function $Z$
(where the number of electrons $N$ and
the projection of spin $s^z$
can fluctuate) and its derivatives are calculated exactly without
taking the thermodynamic limit.
The response
 functions related to electron or hole doping (i.e.
chemical potential $\mu$) or magnetic field $h$ demonstrate
clearly
observable, prominent
peaks paving the way for
 strict definitions of Mott-Hubbard (MH), antiferromagnetic (AF),
spin pseudogaps and related crossover
temperatures~\cite{PRB,JMMM}.

Our exact studies of 4-site clusters indicate a
net electron attraction leading to the formation of bound electron
pairs and possible
 condensation  at finite temperature
for  $U<U_c(T)$
(i.e. suggestive of 
superconductivity)~\cite{PRB,JMMM}.
 This pairing mechanism in the 4-site cluster,  at 1/8 hole doping
 ($\left\langle N\right\rangle\approx 3$)
 away from half filling, exists when the onsite Coulomb interaction
 $U$ is less
 than an analytically obtained critical value,
 $U_c(T=0) = 4.584$
 (in units of the hopping parameter $t$). This critical value, first
 reported in Ref.~\cite{JMMM},
 is temperature
 dependent and can be associated with an energy gap (order parameter)
 which becomes negative below
$U_c(T)$
 implying that it is more
energetically favorable to have a bound pair of
 electrons (or holes) compared to two unpaired
 ones at an optimal chemical potential
 (or doping level) $\mu = \mu_P = 0.658$.
 Above this critical
 value $U_c(T)$,
 there is a Mott-Hubbard like gap that exists when the average
 particle number
 $\left\langle N\right\rangle\approx 3$;
 this gap decreases monotonically as $U$
 decreases and vanishes at
 $U_c(T)$.
 The vanishing of the gap can be
 directly linked to the {\sl onset of pair formation}.
 There is an interval (width) around
 $\mu_P$,
where the pairing phase competes with a phase (having a high
magnetic
 susceptibility) that suppresses pairing at `moderate' temperatures.

\begin{figure} 
\begin{center}
\includegraphics*[width=20pc]{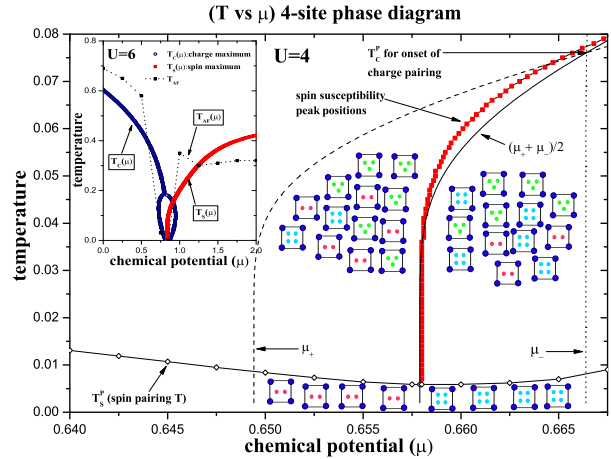}
\hfill
\end{center}
\caption {The
$T$-$\mu$ phase diagram near $\mu_P=0.658$
 ($\left\langle N\right\rangle\approx 3$)
 at $U=4$ for the 4-site cluster.
 The inset shows a
corresponding section (at a different scale) of the
$T$-$\mu$ phase diagram for $U=6$. 
 For $U=4$, note how the paired states condense at
low temperature with a nonzero pair binding energy, while at
higher temperatures, unpaired states begin to appear. This
picture supports the idea that there is inhomogeneous, electronic phase
separation here. When $U$ is higher than $U_c(0)=4.584$,
these inhomogeneities disappear and a Mott-Hubbard like
stable paramagnetic insulating
region results around optimal doping. Note how the (low
temperature) region around optimal doping changes from a
 pairing phase at $U=4$ to a paramagnetic phase at $U=6$ (inset) with
 charge-spin separation as described in the text
and Ref.~\cite{PRB}.}
\label{fig:ph4_6}
\end{figure}

An enlarged view of the
$T$-$\mu$ phase diagram, for the 4-site cluster near $\mu_P$, is shown in
Fig.~\ref{fig:ph4_6}. This exact phase diagram
(at $U=4$)
 in the vicinity of the optimally doped ($N\approx 3$) regime has been
constructed using the ideas described in the text and
Ref.~\cite{PRB}.  The electron pairing temperature,
$T_{c}^{P}$,
identifies the onset of charge pairing. As temperature
is further lowered, spin pairs begin to form at
$T_{s}^{P}$.
At this temperature (with zero
magnetic field),
spin susceptibilities become very weak indicating the
disappearance of the $\left\langle N\right\rangle\approx 3$
states. Below this spin pairing temperature, only paired states
are observed to exist having a
 certain rigidity, so that a nonzero magnetic field or a finite
 temperature is required to break
 the pairs.
From a detailed analysis, it becomes evident that the system is on the
verge of an instability; the paired phase competing with
 a phase that suppresses pairing
 which has a high, zero-field magnetic susceptibility.
 As the temperature is lowered,
 the number of
 $\left\langle N\right\rangle\approx 3$
 (unpaired) clusters begins to decrease
 while a mixture of (paired)
 $\left\langle N\right\rangle\approx 2$ and
 $\left\langle N\right\rangle\approx 4$
 clusters appears.
  Interestingly, the
  critical doping  $\mu_P$
 (which corresponds to a filling
 factor of $1/8$ hole-doping away from half filling),
 where the above pairing
 fluctuations take place when
 $U<U_c(T)$,
  is close to the doping level near which numerous intriguing
 properties have been observed
 in the hole-doped HTSCs.

\begin{figure} 
\begin{center}
\includegraphics*[width=20pc]{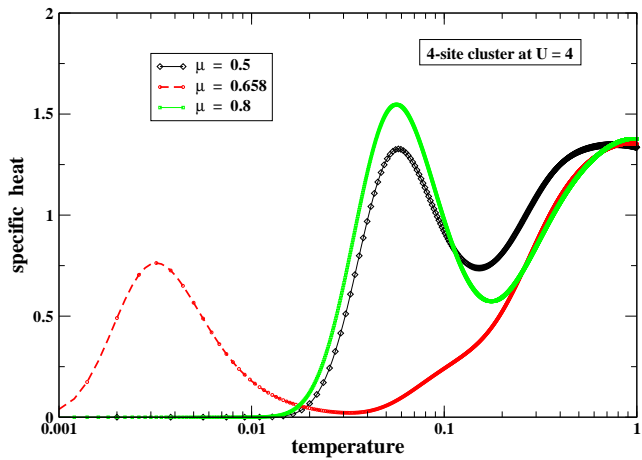}
\end{center}
\caption {Specific heat vs temperature at $U=4$, calculated in the
grand canonical ensemble for the 4-site cluster at several doping
values near the critical doping,
 $\mu_P\approx$ 0.658.
 Note how the low
temperature peak shifts to higher temperatures  when
 the doping is changed from its  
critical value.}
\label{fig:sp_ht}
\end{figure}

\begin{figure} 
\begin{center}
\includegraphics*[width=20pc]{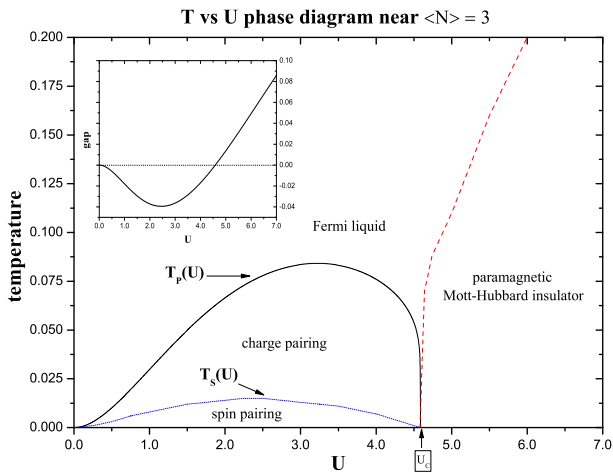}
\hfill
\end{center}
\caption { $T$ vs $U$
phase diagram for the optimally doped
 4-site clusters, based on our exact calculations.
 The inset shows   
 the charge gap
  as a function of $U$ at zero temperature. A negative charge gap
 implies charge pairing.}
\label{fig:ph_tu}
\end{figure}
 Specific heat
 calculations (Fig.~\ref{fig:sp_ht} ),
associated with energy fluctuations,
  also provide further support
 for an electronic
phase change at low temperature.
   There is strong evidence for pair condensation, from specific heat
 calculations shown in the figure.
 As seen in this figure, there is a well separated, low temperature
 peak at $\mu_P=0.658$ (around 40 K, if the hopping
 parameter is set to 1 eV and $U$ to 4 eV).
 This peak, which shifts to higher temperatures when
 the doping level is different from crtical doping,
  is  due to  fluctuations between paired states
 ($\left\langle N\right\rangle = 2$ and
 $\left\langle N\right\rangle = 4$).
 This low temperature peak is in agreement with specific heat
 experiments carried out for the HTSCs~\cite{Tallon}, and is a
 manifestation of the near degeneracy of the states in the neighborhood of
 critical doping $\mu_P$.

Our calculations may also be used to reproduce the
variation of
$T_c (p)$
vs pressure $p$ in the HTSCs,
assuming that $U$ decreases
with pressure~\cite{Chen}. Fig.~\ref{fig:ph_tu} shows condensation
of electron charge below $T\leq T_P(U)$ with bound charge $2e$ and
decoupled spin
 $\left\langle s^z\right\rangle = 1/2$
 for the 4-site cluster.
Below the lower curve $T_s(U)$, the spin degrees are bounded and a
finite applied magnetic field is needed to break them
~\cite{Sebold,PRB}.
Thus below $T\leq T_s(U)$, both the charge and spin
are condensed and the spin degrees can
follow those of charge. The inset in Fig.~\ref{fig:ph_tu} shows
the variation of the charge gap,  $E(2) + E(4) - 2E(3)$, 
 as a function of $U$ where $E(N)$ 
 refers to the canonical energies for $N$ electrons at $T=0$.
  When this gap is negative, pairing is
 favored as discussed in Ref.~\cite{PRB}. 
 In addition, the increase of
 $T_s(U)$ reproduces
 $T_c$ (superconducting transition temperature) {\it vs}
 pressure $p$ in the optimally and nearly
 optimally doped HTSC materials~\cite{Xiao},  indicating a significant
 role of pair binding in enhancing $T_c(p)$.

 Exact results for
 $\left\langle N\right\rangle\approx 3$
 in Fig.~\ref{fig:ph_tu} suggest that
the enhancement of $T_c$ in the
optimally doped HTSCs may be due to an increase of pairing with
decreasing $U$ under pressure rather than an increase of the
pressure-induced hole concentration. Thus it
 appears that the 4-site cluster near
 $\left\langle N\right\rangle\approx 3$ indeed
captures the essential physics of
the electron condensation under pressure.

\begin{figure} 
\begin{center}
\includegraphics*[width=20pc]{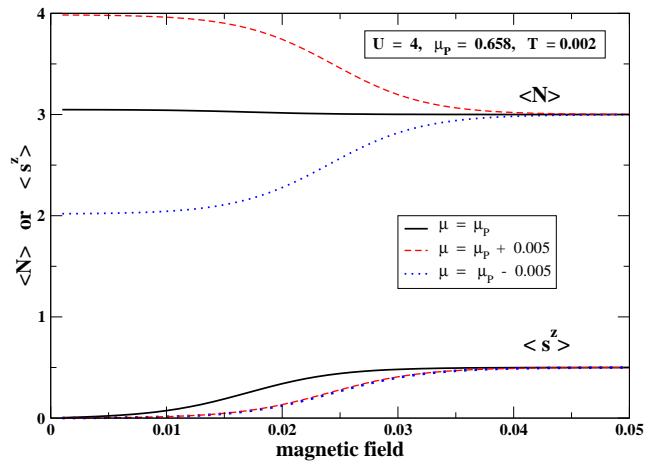}
\hfill
\end{center}
\caption {Variation of electron number
 $\left\langle N\right\rangle$ 
 and magnetization
 $\left\langle s^z\right\rangle$ 
 as a function of external magnetic field for
 several values close to the critical doping $\mu_P = 0.658$ 
at $T=0.002$ and U=4  for the 4-site cluster. Note how
the $\left\langle N\right\rangle = 3$ clusters get stabilized in a
nonzero magnetic field at low temperature. These results support
the recent observation of a dormant magnetic state near optimal doping in
hole-doped La cuprates~\cite{hashini}.}
\label{fig:degree_freedom_1}
\end{figure}

\begin{figure} 
\begin{center}
\includegraphics*[width=20pc]{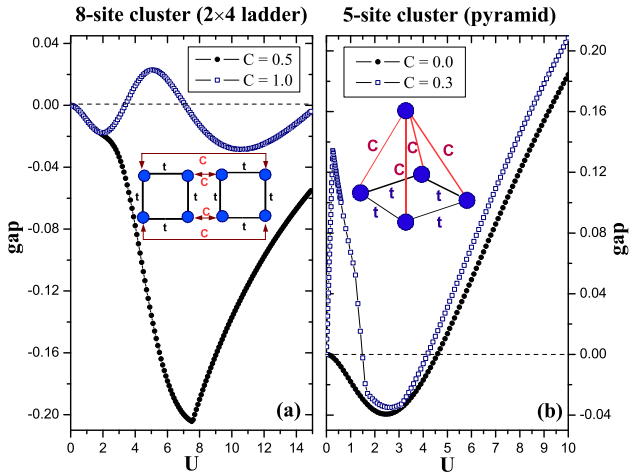}
\hfill
\end{center}
\caption {(a) Charge gaps for the $2{\rm x}4$  cluster at $T=0$ 
 for various couplings $c$ between the squares. (b) Charge
gaps for the 5-site cluster 
illustrate how the coupling $c$ to a 5th site
above the square is responsible for narrowing the window of
$U$ for which pairing exists. In both (a) and (b), the doping level
 is one electron off half-filling and the couplings $t$
 within the squares are set to $1$. Compare these with the inset of
 Fig.~\ref{fig:ph_tu}.}
 \label{fig:gap0204}
\end{figure}

 Another intriguing fact emerging from the exact thermal studies of
the 4-site clusters is the existence of a magnetic state (unpaired
states with $\left\langle N\right\rangle = 3$) with a large
magnetic susceptibility.
 At rather
low temperature $T\le T^{P}_{s}$, this state is 
 dormant.
 However, a small magnetic field or a
 change in chemical
potential can stabilize it  over the paired states
 $\left\langle N\right\rangle\approx 2,4$ as seen from
 Fig.~\ref{fig:degree_freedom_1}
 and the calculated grand canonical probabilities (not shown).
 The variation of the magnetic field mimics the 
 doping to some extent here. Small changes in doping
 (at zero field) can also switch
 the system from one state to
 another with a different
 $\left\langle N\right\rangle$.
 These are useful for understanding
  some recent experimental results reported in Ref~\cite{hashini},
 where a magnetic (and non-superconducting) state has been observed
 near 1/8 hole-doping in La$_{2-x}$Sr$_x$CuO$_{4+y}$. This system is
 said to be on the verge of an instability, surprisingly similar to
 what we observe in these clusters at
 $\left\langle N\right\rangle\approx 3$
 (i.e. near optimal doping away from half-filling).

   In order to monitor the size effects on the properties described above
 for the 4-site cluster, we have carried out a series of
 numerical
 calculations  for
 clusters with different topologies and sizes.
 Fig.~\ref{fig:gap0204}
 illustrates one such set of calculations
 of charge gaps done on a 8-site cluster
 ($2{\rm x}4$ ladder),
 where the hopping term or coupling $c$ between the two squares was allowed
 to be different from the coupling within a given square. The pairing
 fluctuations that are seen for the 4-site cluster exists even for these
ladders near half filling
  ($\left\langle N\right\rangle\approx$  7),
 and  most of the trends observed for the 4-site clusters,
 such as the MH like charge gaps  and
 vanishing of such gaps at critical $U$ values, 
 remain valid here. The fluctuations that occur here at optimal
 doping are among
 the states with  $\left\langle N\right\rangle\approx$ 6, 7 and 8
 electrons.
 Clearly,
 the dormant magnetic state corresponds to
  $\left\langle N\right\rangle\approx$ 7.
 In addition, a 5-site
 pyramid with a square base shows a pairing gap when the coupling $c$ to the
 fifth site (i.e. the site above the square) is weak (up to about
 0.4$t$ where $t$ is the hopping parameter in the plane)
 and disappears above this coupling strength.

 All of the above, from the 4-site and larger cluster calculations, 
 points to a pair binding instability
 near optimal doping  at relatively low temperature.
 Thermal and quantum
fluctuations in the density of holes between the clusters (for
$U<U_c(0)$) make it energetically more favorable to form pairs.
In this case, snapshots of the system at relatively low
temperatures and at a critical
doping level (such as $\mu_P$ in Fig.~\ref{fig:ph4_6})
would reveal phase separation and equal probabilities of finding
in the ensemble of hole-rich or hole-poor clusters only.


In summary, the above results
demonstrate
the importance of the {\it many-body
 interactions}  in microscopic clusters.
 Our exact Hubbard cluster calculations show the existence of
 charge and spin pairing, electronic phase separation, pseudogaps
 and condensation and hence demonstrate a rich variety of
 properties which can be tuned by  doping.
 Furthermore, it is quite surprising to see the number of
  properties that these
 exact clusters share with the HTSCs. This may be, at least in part,
 due to the fact that in all these `bad' metallic high $T_c$
 materials, short-range correlations play a key role.

This research was supported in part by the U.S. Department of
Energy under Contract No. DE-AC02-98CH10886.


\begin{thebibliography}{0}

\bibitem{RVB}
P.~W.~Anderson, Science {\bf 235}, 1196 (1987).
\bibitem{Anderson} P.~W.~Anderson, Adv.~Rev. {\bf 46}, 3 (1997).

\bibitem{Nature}
V. Emery and S. A. Kivelson, Nature (London), {\bf 374}, 434
(1995).
\bibitem{Timusk} T.~Timusk and B.~Statt, Rep.~Prog.~Phys. {\bf 62}, 61
(1999).
\bibitem{Kivelson_Review} S.~A.~Kivelson {\it et al.}, Rev.~Mod.~Phys.
{\bf 75}, 1201 (2003).

\bibitem{Marshall} D.~S.~Marshall {\it et al.}, Phys.~Rev.~Lett.
{\bf 76}, 4841 (1996).

\bibitem{Andrea} Andrea Damascelli, Zahid Hussain,
Zhi-Xun Shen, Rev. Mod. Phys. {\bf 75}, 473 (2003).

\bibitem{Zha} Y. Zha, V. Barzykin and D. Pines, Phys.~Rev. B{\bf 54}, 7561 (1996).

\bibitem{Pastor} F.
Lopez-Urias and G. M. Pastor, Phys.~Rev. B {\bf 59}, 5223 (1999).
\bibitem{Sebold} A.~N.~Kocharian and Joel~H.~Sebold, Phys.~Rev. B {\bf 53}, 12804 (1996).
\bibitem{scalettar} R.~M. Fye, M.~J. Martins, and R.~T. Scalettar,
Phys.~Rev. B {\bf 42}, R6809 (1990).
\bibitem{white} Steven R.
White, Sudip Chakravarty, Martin P. Gelfand, and Steven A.
Kivelson, Phys.~Rev. B {\bf 45}, 5062 (1992).
\bibitem{Shiba}  H. Shiba and P. A.
Pincus, Phys. Rev. B {\bf 5}, 1966 (1972).
\bibitem{schumann} R.~Schumann, Ann.~Phys. {\bf 11}, 49 (2002).


\bibitem{PRB} A.~N.~Kocharian, G.~W.~Fernando, K. Palandage and
J.~W.~Davenport, Phys.~Rev. B {\bf 74}, 024511 (2006).

\bibitem{Tranquada} J. M. Tranquada, B. J. Sternlieb, J. D. Axe,
   Y. Nakamura, S. Uchida, Nature, {\bf 375}, 561 (1995).

\bibitem{Tallon} G. V. M. Williams, J.~L.~Tallon and J.~W.~Loram,
Phys.~Rev. B {\bf 58}, 15053 (1998).

\bibitem{Xiao}
Xiao-Jia Chen, Viktor V. Struzhkin, Russell J. Hemley, Ho-kwang
Mao, and Chris Kendziora, Phys.~Rev. B{\bf 70}, 214502 (2004).

\bibitem{hashini} H. E. Mohottala, B. O. Wells, J. I. Budnick,
 W. A. Hines, C. Niedermayer,
L. Udby, C. Bernard, A. R. Moodenbaugh and Fang-Cheng Chou, Nature
(London), {\bf 5}, 377 (2006).

\bibitem{nmr} Y. Itoh, M. Matsumara and H. Yamagata, J. Phys. Soc. Jpn,
{\bf 66}, 3383 (1997).

\bibitem{JMMM} A.~N.~Kocharian, G.~W.~Fernando, K. Palandage and
J.~W.~Davenport, J.~Mag.~Mag.~Mater., {\bf 300}, e585 (2006).

\bibitem{Chen} X. J. Chen, H. Q. Lin, C. D. Gong, Phys. Rev. Lett.
{\bf 85}, 2180 (2000).


\end{thebibliography}
\end{document}